# Value, Variety and Viability: Designing For Co-creation in a Complex System of Direct and Indirect (goods) Service Value Proposition

**Irene Ng**
Service Systems Research Group
Warwick Manufacturing Group
University of Warwick
irene.ng@wmg.warwick.ac.uk

**Gerard Briscoe**
Systems Research Group
Computer Laboratory
University of Cambridge
gerard.briscoe@cl.cam.ac.uk

**Abstract**

**Purpose**
While service-dominant logic proposes that all "Goods are a distribution mechanism for service provision" (FP3), there is a need to understand when and why a firm would utilise direct or indirect (goods) service provision, and the interactions between them, to co-create value with the customer.

**Methodology/Approach**
Three longitudinal case studies in B2B equipment-based 'complex service' systems were analysed to gain an understanding of customers' co-creation activities to achieve outcomes.

**Findings**
We found the nature of value, degree of contextual variety and the firm's legacy viability to be viability threats. To counter this, the firm uses (a) Direct Service Provision for Scalability and Replicability, (b) Indirect Service Provision for variety absorption and co-creating emotional value and customer experience and (c) designing direct and indirect provision for Scalability and Absorptive Resources of the customer

**Research Implications**
The co-creation of complex multidimensional value could be delivered through different value propositions of the firm. The research proposes a value-centric way of understanding the interactions between direct and indirect service provision in the design of the firm's value proposition and proposes a viable systems approach towards reorganising the firm.

**Practical Implications**
The study provides a way for managers to understand the effectiveness (rather than efficiency) of the firm in co-creating value as a major issue in the design of complex socio-technical systems.

**Originality/Value**
Goods are often designed within the domain of engineering and product design, often placing human activity as a supporting role to the equipment. Through an SDLogic lens, this study considers the design of both equipment and human activity on an equal footing for value co-creation with the customer, and it yielded interesting results on when direct provisioning (goods) should be redesigned, considering all activities equally.

**Keywords:** Complex Service Systems, Service Dominant Logic, Customer Experience, Design, Value Co-creation

# 1 Introduction

Managing organisational performance has become increasingly complex as firms 'add value' predominantly through the provision of services (Baines, 2007). This challenge is widespread, from manufacturing to information systems. In manufacturing this is commonly known as the phenomenon of 'servitization'; where there has been consideration on motives, benefits and feasibility of servitization as a competitive strategy (Vandermerwe, 1988). Within Information Systems (IS) the firm is traditionally seen as the provider of service, but this is increasingly being challenged, also under the conceptualisation of service as proposed through the Service-Dominant (S-D) Logic (Vargo and Lusch, 2004, 2008). Under S-D Logic a firm should be reorganised to offer direct or indirect (goods) service provision, and it should understand how to manage the interactions between them to co-create value with the customer. Our paper shows that the following constructs are critical in designing for this:

First, In the understanding of **value**, contemporary literature has moved the discussion away from the traditional understanding of value as value-in-exchange to the concept of value-in-use, where the customer would realise the firm's value proposition in consumption and experience.

Second, understanding this value-in-use requires fully appreciating the context in which value is created, more so the contextual **variety** in which services are consumed, because greater variety in use has considerable implications for service provisioning.

Third, firms need to maintain **viability** from a viable systems perspective when co-creating value with their customers, and achieve value-in-use irrespective of the contextual variety. The viable systems approach provides a model for the organisational structure of the firm as it makes the transition from being a manufacturer to a system of achieving value-in-use in co-creation with the customer.

Our paper analyses three longitudinal case studies of manufacturers moving into outcome-based service provision over three years. We propose that the understanding of value-in-use, contextual variety, and a system's perspective of viability are the **three core principles** for designing an organisation that is able to co-create value with customers through both direct and indirect service provision.

We found the nature of emotional value to be co-created i.e. the customer experience, the degree of contextual variety and firm's 'legacy' viability as the three challenges of achieving co-creation as it threatens the viability of the firm. To counter the viability threat, the firm uses (a) Direct Service Provision for Scalability and Replicability of value proposition, (b) Indirect Service Provision for variety absorption and co-creating emotional value (customer experience) and (c) Scalability and Absorptive Resources of the customer that would impact on the firm's provision. Overall, the firms came to the realisation that an asset was not a 'sacred cow' and that the better it was at absorbing contextual variety of use, the less it would depend on human capability and the easier it would be to scale and replicate across contracts. Furthermore, our study suggests that organisations structured around manufacturing require a re-evaluation of their operational elements and viability when they transform into a full-service organisation. We argue for a transformation in the customer relationship to help realise the value proposition that firms offer. Specifically, we propose a viable systems approach for the inclusion of customer activities within the firm's boundaries of management and operation for value co-creation, and our paper argues how this could be achieved while maintaining viability.

The remainder of the paper is organised as follows. A literature review considering the theoretical links between value, variety and viability in a complex service system is followed by the methodology which describes the longitudinal case studies of manufacturers who have contracted based upon outcomes. The findings from these case studies are used to address the research question. We then discuss the extension of the S-D Logic approach towards organising the firm through a viable systems approach, before concluding with the managerial implications on this new way of configuring the organisation.

## 2   Literature Review: Theoretical Links Between Value, Variety and Viability in a Complex Service System

### 2.1   Value

Value has been subjected to much discussion and debate over the centuries. Despite a common etymological origin, the term has evolved into two distinct meanings. The first describes value as 'goodness' determined by an individual personally and culturally, and in an ethical sense. Such values are held most dear by an individual and govern what the individual does and becomes (e.g. Weber, 1909). The second meaning, and the subject of this paper, also describes value as 'goodness' but in its description of something; be it a person, an idea, a product, an activity or anything else. The study of the latter meaning of value as described above has a deep mathematical and philosophical history, rooted in the discipline of axiology. Axiology, the philosophical study of value, concerns itself with the analysis of value, its frameworks and the evaluation of what is 'valuable', or with the assignment of value to items, to properties or to states (Bengtsson, 2004).

In *The Republic* (360 BCE), Plato proposes the notion of intrinsic and instrumental (extrinsic) value. In this proposition Plato suggests that items with extrinsic value are good to have, as they are instrumental to achieve or obtain something else that is good. Whereas an item that is intrinsically of value is good to have for itself. To Plato, the two are not mutually exclusive and products could exhibit both intrinsic and extrinsic properties.

Scholars developing academic thought around value have suggested different ways of classifying extrinsic or intrinsic value, albeit with different degrees of robustness. Mattsson (1992) for example, suggests that intrinsic value is analogous to the emotional dimension of value, whilst extrinsic value could have practical and logical dimensions. Consequently, a chair has the practical value of a 'seat' and has the logical value of 'width, size or height' but could also have some emotional value of being 'great-grandpa's chair'. Hartmann et al (1967) introduce a further concept of value – that of systemic value – where the characteristics of the thing that is good has finite properties defined by a system, or the norm. Thus, a chair is only good if it can seat a person without falling over, since all good chairs share the same property.

The idea of extrinsic value has also been developed by (Marx, 1867), where not only is the item purposeful, but its value can only be realised in context. Marx described it as "value only in use, and is realised only in the process of consumption" . Such discussions have one commonality in their descriptions. As proposed by Moore and Baldwin (1993) and Hartmann et al (1967), the commonality among descriptions is that value can only be determined and evaluated by the perceiver.

While intuitively logical, Hartman et al (1967), Haglund (1988), Mattsson (1992) and other scholarly extensions of Hartman's (1973) work contrast with the phenomenological concept of value. Instead of seeing objects as having some properties emanating from them, Husserl et al (1973) proposed a phenomenological way of looking at objects by suggesting that individuals, in their own way re-constitute such objects such that the object ceases to be something simply "external", but become part of the individual's group of perceptual purpose. Phenomenological value therefore regards objects as inherently conceived in the experience of it i.e. in the interaction or relationship between the item and the perceiver. Therefore, understanding phenomenological value is the understanding of how to identify the invariant characteristics of items and how such characteristics have a role in the way individuals perceive the reality. In so far as 'great-grandpa's chair' is concerned, it can only be of value as experienced (mentally or physically) by the valuer within his/her consciousness (Husserl, 1939 [1973]). Such a view of value differs from Hartman's in the sense that it is a systems theoretic view of value. The value is emergent and experienced between object and subject, as compared to Hartman's reductionist view of value, where value lies within an object and subject. A reductionist view of value in objects and perceivers may not give sufficient weight to the interactions between them, interactions that would serve the individual's contextual purpose. Such a value concept is consistent with Vargo and Lusch (2004, 2008] who consider use-value to be phenomenologically determined, that it is "uniquely and contextually interpreted" (Vargo and Lusch 2008, p.4]. In addition, literature in consumer culture theory has long since discussed the notion of consumption as experiential within a social and cultural phenomenon (Belk and Sherry, 2007; Arnould and Thompson, 2005].

This discussion of value is important as it resonates with scholars who have described value as that which an individual derives from an offering because of the individual's ability to co-create that value with the offering to achieve his/her outcomes (Vargo and Lusch 2004, 2008; Tuli *et al*, 2007). So, such value co-creation occurs through a process of an individual integrating his/her resources with the offering to achieve value. The co-creation of value is central S-D Logic (Vargo and Lusch, 2004, 2008), which conceptualises service as the co-creation of value between the individual and the firm through an integration of resources accessible to both parties. It has therefore been proposed [Vargo and Lusch, 2004, 2008) that firms do not provide value, but instead value propositions realised through co-creation interactions with the individual to achieve his/her goals. As we have discussed above, value can include an emotional dimension (Mattsson, 1992), and therefore so can that of value co-created in use. For example, using an online service to share photos can strengthen one's emotional bond with friends and family.

This co-created value-in-use begins with the enactment of value propositions, with product offerings that are value unrealised until the individual customer realises it through co-creation, thus gaining the benefits (Ballantyne and Varey, 2006). So, understanding co-creation requires understanding customer consumption processes as the customer determines value-in-use through consumption and confirmation (Ballantyne and Varey, 2006). As such, achieving value-in-use through co-creation has received considerable attention (Payne *et al*, 2008; Grönroos and Ravald, 2010; Sandström *et al*, 2008; Heinonen and Strandvik, 2009], and most have acknowledged that value-in-use is achieved in context, while some have proposed that value-in-use is synonymous with value-in-context (Vargo and Lusch, 2008). As value-in-use is co-created in context, and because the context is not necessarily known beforehand, there is the potential for new experimental use to occur. For example, becoming unexpectedly lost on holiday, such that the 'maps' feature (application) on a smartphone becomes invaluable.

With that understanding of co-creation, we now explain the concept of contextual variety.

## 2.3 Variety

Given that value is created in a use situation, contextual conditions of that situation could affect the co-creation [for literature on situational and contextual value, see Beverland *et al* (2004), Flint *et al* (2002), Lemon *et al* (2002), Lapierre *et al* (2008)]. Palmetier (2008) states that contextual variables may arise from changes in the physical environment, originating either from the provider and/or from the customer themselves. In any use of an offering, there could be a number of contextual factors affecting value creation, and such contextual factors will result in contextual variety in the way value is co-created, even by the same individual. This is consistent with a systems perspective, where variety is the measure of the number of different states in a system. Variety is the measure of complexity as it counts the number of possible states of a system. *Contextual* variety as we describe here, is the number of different states in a system caused by different contexts of use i.e. the sets of value possibilities that an offering could be experienced in co-creation with the individual user.

Contextual Variety also has some theoretical foundations within Economics, particularly in state-dependent utility, an economic term describing how the state of the world affects how well individuals are able to enjoy the consumption or utility of a product (Cook and Graham, 1977; Karni, 1983). State-dependent utility has had a following in pricing literature (e.g. Xie and Shugan, 2001, Shugan and Xie, 2000, Png, 1989), where the decision to buy or the willingness to pay for a product could be influenced by the state- dependent nature of the utility the individual derives from the product at a specific time in the future. However, state dependency does not merely affect exchange value or the price the firm can charge. It also has an impact on value-in-use, in so far as the context of use might be a result of the state of the world that could disrupt or enhance the individual's value co-creation and use of the firm's offering in context. Ng (2008) describes state dependencies as contingent factors which could arise from changing firm, customer or environmental states in use situations. Theoretically, we can therefore conceptualise Contextual Variety as the varied nature of the value created in use situations due to contingencies caused by the state of the world (context). From an S-D Logic perspective, this does not contradict the notion of value being always *uniquely* experienced, as we argue that the uniqueness in individuals' value is not conceptually the same as a pattern of use. For example, the use of a kettle may be uniquely experienced and valued on a day-to-day basis, but its Contextual Variety could be low if the kettle it is always on the kitchen worktop or very high if it is brought to different places daily (e.g. a boat, the office etc.). The degree of contextual variety is therefore related to what an offering was designed to **do** i.e. its most likely purpose of use within most likely contexts. It is when contexts begin to change more rapidly and not according to normal expected contexts of use that the degree of contextual variety increases. For example, an LCD TV is designed to provide entertainment at a fixed physical location. However, if it suddenly finds itself useful as a screen for presentation purposes, it could be moved around and certain parts of the TV may find themselves less able to take the stress from a change of use e.g. the pedestal on which it sits may break from being shaken about too often. Thus, a high degree of contextual variety is an increase in contexts (states) that deviate from the most likely contexts (states) of use for which the offering was originally designed.

In manufacturing literature, requirements analysis or requirements engineering is a critical component of product design. These are tasks that determine the needs or conditions of product use, taking into account conflicting requirements of various stakeholders such as end-users, transporters etc. (Boothroyd, 1994). Yet, research has shown that requirements gathering may not be able to understand, exhaustively, all the sets of possibilities surrounding customer requirements for the use of the asset (Potts and Hsi, 1997). In addition, even if the firm recognises that there are *n* sets of possibilities in which an asset could be used, it may not be feasible or viable to design the asset for every possibility. For example, it would be completely unfeasible to manufacture an aircraft or an engine that is be able to fly through an ash cloud 365 days in a year. That implies that every product is a manifestation of tradeoffs between different sets of possibilities in contextual use and the firm has to acknowledge that there will be some contextual variety, where such variety arises from the set of possibilities not taken into account, or not deemed to be feasible for the design and manufacture of the product.

The above discussion of value and variety now brings us to a third perspective, that of viability.

## 2.2   Viability

Stafford Beer (1979, 1981, 1985) introduced the Viable Systems Model (VSM) as that which describes the necessary conditions for viability. Viability is defined as *the ability to maintain an independent* *existence within a specified environment*. In business, a firm that is viable is able to obtain funding or revenues for its offerings such that it is above the cost of delivering such offerings. The management structure of the firm exists to support the firm to continue the process of profiting from its offering, without which it would quickly become non-viable. Viable systems approach has been touted as one of the more robust approaches to describe viability, and can be applied to governments, firms, biological systems and other ecosystems. However, as Beer (1985) would qualify, the analysis of an organisation using the viable systems approach is not to determine the facts but to establish a set of conventions that are neither true nor false, but are more or less useful.

The viable systems approach suggests that there are five systems that are necessary to ensure viability. The five systems are illustrated in Table 1.

System 1, shown in Figure 1, is where the firm operates within an environment, depicted by a grey oval form. This system has to deliver what it has been designed to do, despite changes in the environment, so it must have the capacity to adapt in order to be able to cope and return the entity to stability. Beer refers to the fundamental operations within a viable system as its System 1, made up of all the operations which do the things that justify the existence of the system. This includes the management of these operations but excludes senior management, which is considered as a set of services to System 1. Without System 1, there would be no reason for the firm to exist. A firm's environment consists of its customers, suppliers and regulators, all of which could experience perturbation which in turn could disrupt the firm's core System 1 operations. Collectively, Systems 5/4/3 represent the meta system (future planning) and Systems 1/2/3 represent the current system (present planning) with System 3 as the key controlling bridge between the activities of Systems 1/2 and the management of Systems 4/5. To achieve *homeostasis*, i.e. the property of a system that regulates its internal environment and tends to maintain a stable, constant condition, it must be enabled by *resources*, and governed by *management* (Golinelli, 2010). There are three main aggregate homeostats in VSM. The first is the homeostat in System 1 that stabilises the operations of the firm with its markets along the horizontal axis (first axiom of management). The second is the homeostat 3/4 maintaining System 3's coordination of the present with System 4's focus on the future (second axiom of management). The third is the homeostat that balances the horizontal variety between the System 1s and their environment and the vertical variety from Systems 1 to 5 (third axiom of management). These three homeostats achieve stability in the firm to ensure its continued viability. It is important to note that the system in focus has to have a purpose. "Without a purpose, it is impossible to define a systems boundary". "An essential basis for identifying and organising a system structure is to have a sharply and properly defined purpose" (Forrester, 1968 as quoted in Richardson and Pugh, 1981). The boundary of the system is an imaginary line separating what is inside from what is outside, for modelling purposes.

The creation of system boundaries has many implications, including the potential for recursive behaviour. Recursion is essentially the process that an activity (procedure) goes through when one of the steps of the activity involves invoking the activity itself (often with a different set of parameters). This of course risks an endless loop, but recursion can be defined such that in certain cases (sets of parameters) the activity completes, no longer calling itself. Within computer science and mathematics it is more formally defined as a function (routine) that calls to itself, where a recursive function is stopped by one or more base cases for which the result is predefined, and so does not require the function to call itself to determine the results as it might have done previously (Cutland, 1980).

## 3     Value, Variety and Viability: Designing a Viable System for Value Co-creation

The focus of this paper is to analyse a firm's System 1 operations as it moves from being a manufacturing firm to becoming a service-oriented firm, where the value proposition changes from the manufacturing of an asset to the co-creation of outcomes in a combination of assets, as well as human activities. Such a move transforms System 1's operational purpose from that of 'making something' to 'delivering solutions or outcomes'. The latter operations often result in the System 1 operation being a complex delivery system of people, processes, technologies and equipment and there is less understanding of what framework could be used to inform the configuration of System 1 resources to achieve viability, whilst ensuring outcomes are achieved. Clearly, no one can make a complex system less complex. However, as Beer professes, "By finding invariances that underly viability, is to make all of it susceptive to uniform description" (Beer, 1985, p.15). The notion of an *invariant* i.e. a factor in a complex situation that is unaffected by all the changes surrounding it, is explored, and the purpose of this paper is to derive an invariant framework required for a manufacturer to achieve service transformation and achieve viability to deliver outcomes to the customer.

As a firm moves from manufacturing an asset to be sold to achieving outcomes, it immediately inherits the problem of contextual variety, as discussed earlier and also pointed out by (Ng *et al* 2009, 2011]. It is one thing to deliver an asset to customers and leave them to realise the value in their own space and time, but quite another for the firm to promise customers that their outcomes can be achieved across the customers' varied contexts. Yet, contracts such as Rolls-Royce power-by-the-hour®, or a bank of flying hours of a fastjet in ATTAC (Ng *et al*, 2010) are promising just that, which immediately result in the firm facing a high degree of contextual variety to achieve value-in-use (Ng *et al*, 2011). This is subject to the Law of Requisite Variety which originates from the field of cybernetics, control and systems theory (Ashby, 1956); this essentially states that in active regulation only variety can destroy variety (Ashby, 1969). In other words, the more complex and variable a system becomes, the more flexibility and variety is required to manage those changes. This leads to the somewhat counter-intuitive observation that the regulator must have a sufficiently large variety of actions to ensure a sufficiently small variety of outcomes. Furthermore, it has important implications for practical situations; since the variety of perturbations a system can potentially be confronted with is unlimited, we should always try to maximise its internal variety to be optimally prepared for any foreseeable or unforeseeable contingency (Heylighen and Joslyn, 2001). Naturally, this has implications for systems of all types, including organisations, economics, families, interpersonal relationships and mental processes, except for quantum systems not subject to traditional causality. Therefore, it has been applied to many fields, from game theory (Ashby, 1958) to politics (Jessop, 2003).

The Law of Requisite Variety clearly has implications within the Viable Systems Model, i.e. for a system to remain viable, variety must be managed. The Law of Requisite Variety was restated as only variety can absorb variety (Beer, 1979), meaning that to manage or control something effectively, you need to be able to at least equal the variety that the thing itself exhibits. Stated more simply, the logarithmic measure of variety represents the minimum number of choices needed to resolve uncertainty, which is used to allocate the management resources necessary to maintain process viability (Beer, 1981). Alternatively, firms can design approaches to amplify or attenuate variety that enhances the ability to manage, and so increase performance regardless of the contextual variety being operated within. Different parts of a firm as a system need to treat variety differently and may need to amplify their variety (e.g. responses) rather than attenuate them (Godsiff, 2000), depending primarily on their level of direction interaction with the customer in value co-creation.

Whether the firm seeks to attenuate or amplify variety, the stability of System 1 is dependent on resources deployed to achieve homeostasis and improve viability. Current literature does not provide any answers towards the resource configuration required within System 1 to successfully co-create value with the customer, where resources to co-create value are a combination of assets (equipment or goods) and human activities (people and processes). Indeed, most literature refer to the notion of 'servitization' as simply adding on service features (human activities) that relate to the core tangible asset to create additional exchange value, and consequently, boost revenues and the bottom line. There are very few literature that offer a framework to understand how value could be co-created to achieve outcomes when the value proposition is a combination of assets and people, within a system of processes and in combination with the customer activities.

S-D Logic (Vargo and Lusch, 2004, 2008) suggests a way forward. It proposes that "Goods are a distribution mechanism for service provision" (FP3) and that all offerings are service. While conceptually, it can be regarded that all offerings aim to deliver outcomes, it can be argued that the outcome achieved through an 'indirect service provision' (product) requires more customer resource to realise than an outcome made possible through a firm's direct service activities, a point acknowledged by (Vargo and Akaka, 2009). In addition, the capability to achieve the same outcomes whether through direct or indirect service provision requires a different set of capabilities from the firm. Neely (2008) provides empirical evidence that despite increasing cases of organisations throughout the world attempting to deliver solutions by adding service activities, such firms often generate lower profits as a percentage of revenues compared to pure manufacturing firms. Neely (2008) attributes this to the organisational challenges of the inevitable changes to value propositions that such a change in capability entails. This is echoed throughout discussions in academic literature, as many authors continue to highlight the need to explore the operational implications of transitioning from manufacturing to service (e.g. Pawar *et al.*, 2009; Johnstone *et al.*, 2009; MacDonald *et al.*, 2009; Oliva and Kallenberg, 2003]). Not only do they recognise the need to explore the operational elements of this transformation but they acknowledge the need to do so with a customer orientation (Johnstone *et al.*, 2009), and many look to the S-D Logic [Vargo and Lusch, 2004, 2008] as a lens through which this exploration could be possible [Pawar *et al.*, 2009, Macdonald *et al.*, 2009]. S-D Logic considers value co-creation as a process involving the integration of resources. Yet the resources for co-creation by the firm delivering an indirect provision, which in turn specify the capability of the firm, is clearly different from the resources for the same firm delivering service activities directly. From a viable systems perspective, if the resources to specify the core transformation of System 1 begin to change, creating instability, and the management of System 1 fails to regulate to achieve homeostasis, the firm could quite quickly find itself becoming non-viable as evidenced by firms attempting to 'servitize'.

Consequently, in a firm's transition from a manufacturer to an outcome-driven service provider, we are interested to discover the threats to viability and the drivers to direct or indirect service provision to ensure continued viability even while value, together with its high contextual variety, is being co-created with the customer. This is the research question we seek to answer.

## 4   Methodology

We consider three longitudinal case studies of manufacturers who have contracted based upon outcomes. Case study research is useful when the aim of research is to answer "how" and "why" questions (Yin, 2003). The three contracts analysed were awarded to three different organisations. All three were awarded for the service of equipment they had originally sold to the customer. However, unlike the conventional equipment-based service contracts where the firms are paid based on activities, repairs or spare parts used, the contracts were awarded on the basis of the *availability* of the equipment. The delivery of these contracts serves as an exemplar for complex service systems where both parties are focused on achieving outcomes i.e. flying hours of a fastjet, or missile availability and engine hours in the air; the value is co-produced with the customer (to achieve the outcomes); and the customer co-creates value with the firm through the *use* of the equipment. These service contracts were operating under complex relationships between clients and service providers and therefore relied heavily on both indirect service provision (e.g. tangible equipment) and direct service provision (e.g. knowledge and relationships through human resources) to deliver the outcome of the contract, often through complex supply chain management. As the service systems grow into the maturing phase, they become more complex. As a result, standardisation, automation, and commoditisation were needed to ensure some efficiency.

Each case study data was obtained through qualitative interviews, participant observations and company internal documents. There are a number of different methods to be used in qualitative research such as observation, analysis of texts and documents, interviews, and recording and transcribing (Dooley, 2001). The logic behind using multiple methods is to secure an in-depth understanding of the case.

A total of 50 in-depth interviews were conducted with stakeholders from the firm and the customer over three years, to obtain a longitudinal understanding of the phenomenon. These interviews were audio recorded and subsequently transcribed, coded and categorised.

## 5      Findings

We found the **nature of emotional value to be co-created i.e. the customer experience**, the **degree of contextual variety** and **firm's 'legacy' viability** threatens the viability of the firm. To counter the viability threat, the firm uses **(a) Direct Service Provision for Scalability and Replicability of value proposition, (b) Indirect Service Provision for variety absorption and co-creating emotional value and (c) Scalability and Absorptive Resources of the customer as a influential factor for its direct/indirect provisioning.**

### 5.1     Threats to Viability

#### 5.1.1  Nature of Emotional Value to be Co-created (Customer Experience)

First, the nature of value to be co-created has an impact on the type of resources used in System 1. In all three cases, we found that the value consists of not only practical and logical value (labelled jointly as *functional value*) but also emotional value, in the form of the *experience*. In each of the cases, it was not only the functional value that was important to the customer, but the customer's perception had to be transformed into one that believes outcomes were achieved or achievable. In other words, System 1 not only had to transform materials and equipment to achieve the outcomes; the customer also had to be convinced that the process of doing so was culturally and adequately aligned with the needs of the customer organisation. This meant that previously, when the organisation had only to deliver an asset, System 1 was all about resources for transforming materials and equipment in a factory setting and handing it over to the customer, an indirect service provision. Yet, when the value to be co-created was outcome-based, customer perception of the experience became an important element of that value. The customer became concerned with both the process as well as the achievement of the outcomes, and the firm had to engage with the customer in a different manner and through different resources to ensure the perceptions/experiences were achieved. This was often achieved through relationships:

> *"I don't think we put enough spending into how much relationship is worth as a business. We tend to focus heavily on the things that you can touch and feel like erm somebody can write you a process or a procedure but it's the softer issues that make these things work the softer skills, the you know the way in which people interact, the way in which* *we operate with our customer once we are on his [site]. You know they are the things that really grease the wheels….that's the glue that makes all this work."*

*Proposition 1: In co-creating value for customer experience and emotional outcomes, System 1 for the firm has to include the transformation of the customer to ensure viability*

### 5.1.2 Degree of Contextual Variety

Second, the degree of contextual variety also had an impact on what resources were used in System 1. We found that contextual variety arises not merely from the context of usage, but in the moral hazard from equipment use when there is no sense of ownership. As one respondent puts it:

*" … it's like a car isn't it, you-know? I drive my car and abuse my car, whereas my partner looks after her car, so that gives different demands on the garage. …..If they don't do that in a logical way, following the process that's outlined in the manual – the data that we get back that we need to analyse to try and reduce [problems] on the [asset] and reduce the number of faults on the [asset] is flawed."*

The variety of use became a serious issue as contracts require constant amendment to accommodate increasing sets of possibilities:

*"….The other thing of course is the contract doesn't stay the same, its constantly being changed and then the [outcomes] have changed they are going to want to give you extra work or extra scope so more and more things are coming into the contract and we go oh this is an amendment is that a purely fixed amendment is it variable is a mixture is it, so the baseline changes constantly as we move forward"*

Our study found that contextual variety threatens viability in two ways. The first threat is from the firm being unable to absorb variety. This means that System 1 has not got the requisite variety to absorb contextual variety from use, and implies that the customer may be unhappy due to the firm's inability to accommodate certain contexts of use. This inflexibility threatens the long-term viability of the firm as it struggles to meet customer expectations in a timely manner, and it may find itself losing the customer as a result of that failure. The seecond threat is from absorbing too much variety which disrupts the system, challenging homeostatis. We found that when the contextual variety of use is high, the firm amplifies its variety through greater responses, and System 1 suffers the strain as inadequate resources are provided to stabilise the system.

*Proposition 2: In co-creating value for outcomes, the firm has to balance the attenuation and amplification of internal responses to match contextual variety to ensure viability*

### 5.1.3 'Legacy' Viability of the Firm

Our study found that when System 1 was operating purely as a manufacturer, it did not have to manage much variety. The firm's established viability was based on a transfer of asset ownership and when called upon, undertake maintenance and service activities, relegating the variety issue to a scheduling problem. However, when the firm is tasked to co-create for outcomes, it has to take responsibility for the outcomes within the customers' use situations which results in the firm having to take proactive initiatives that are uncertain and where the absorption of variety may require different resources. It also meant that the transfer of responsibility requires the firm to be involved in customer contexts and use situations so as to obtain the benefit of reduced costs and reduced variety. Yet the quote below shows how this threatens the established system and challenges the mindset:

> *"when I report back into mothership they would say, 'why are you worried about …the user? That's not the contract – you've just got to deliver the [outcome]'. And I'm saying, 'well hang on a minute…….why wouldn't you get closer to them? Because, in most cases, it creates a win-win situation where you're involved in terms of what the customer finally gets and, in financial terms, we gain anyway……but I'm struggling to get the back-end of the company to get that?"*

Since the asset is now the responsibility of the manufacturer to achieve outcomes, the co-creation activity no longer interacts in the same way as when the asset was the responsibility of the customer. Yet, System 3 could be controlling Systems 1/2 in a 'legacy' manner, while Systems 1/2 are struggling to cope with a different kind of variety entering the system. This leads to an imbalance:

> *"I've got somebody sat in the back office at ….. who's just got it in his tray, having a cup of tea and thinking in weeks, months and years, when I'm trying to think in seconds, minutes and hours ….So that means back office needs to change the way they're organised and the way they work and what they're roles and responsibilities are and, in some cases, their capability as well."*

*Proposition 3: In ensuring viability, the firm has to ensure that resources allocated to Systems 1/2 are in line with Systems 1/2 key operational elements and not legacy operational elements*

Our findings suggest that that the choice between indirect and direct provision interacted severely, and there is tension between resources for scalability and replicability (assets and scalable processes) and resources for variety absorption (autonomy, empowerment and human skills) to deliver outcomes. They also show that the choices of direct and indirect provisions improved the viability of the firm in different ways.

## 5.2 Ensuring Viability in Service Transformation

### 5.2.1 Indirect Service Provision for Scalability and Replicability of The Value Proposition

Our findings suggest that when firms were manufacturers, their viability came from production, which could be grown or scaled as long as the order books kept filling up. In co-creating outcomes however, firms became increasingly challenged by the difficulty in scaling or replicating for growth due to embedded human capability.

> *"…and service thing is not easy with this new model…we could get a different person and it won't turn out the same……and then there so many changes that you can't really design anything …the customer wants different things, solve different problems … there's a fire fighting mentality…"*

Our findings show that high indirect service provision within a firm's outcome-based value proposition delivered low margins on a contract for two reasons. First, it makes *the system* less replicable since embedded human capability, particularly when skills and knowledge form a valuable resource for the system, is not as easily transferable to other employees as assets. This then results in slower growth for the firm since the firm's systemic capability to deliver outcomes takes a longer time to acquire. Second, human resource component within a system makes the system less able to scale. Whilst an asset could be scaled up by increasing production lines and/or improving manufacturing capacity, complex service systems of direct and indirect provision are less easily scaled, resulting in investment or costs for a small project that could be similar for a big project. Economies of scale are therefore harder to achieve in outcome-based environments.

To counter the above challenge, the firms in our study found that firms became willing to change indirect service provision to achieve outcomes that could be more scalable and replicable, modifying the asset through redesign or incorporating technology insertions:

*"I think we're achieving better outcomes with the current equipment because we're starting to collect more [electronic health monitoring] data about what's happening; we're starting to have different discussions with the customer about what's happening so we can actually get a better understanding of what's happening and look for failures, or signs of failures happening before they actually fail."*

*Proposition 4: Redesigning and modifying indirect service provision ensures viability through scalability and replicability*

**5.2.2 Direct Service Provision for Variety Absorption and Co-creating Emotional Value and Experience**

Conversely, our study found that the use of direct service provision was essential to absorb contextual variety.

*"You then see that he can then use those relationships to either just sort of oil the wheels altogether speed things up or he could have a conversation say with the [customer employee] ……. he would talk to [person] and [person] would go and do it and at the end of the day the [customer employee] work for him so there is all that sort of complexity of relationship building and then you just know you are going to get benefit from that but things happen, things are much easier, things get smoothed through that could otherwise could become an huge issue."*

The resources used to absorb the impact of variety into the firm were often human in two ways. First, in direct engagement with the customer, the firm would try to ensure low contextual variety by monitoring and engaging the customer on use behaviour:

*"So what it's driven us to do is start to focus more on managing [problems] and to do that we need to get closer to the user…. What are you doing with it? How are you [using] it … Erm, how are you looking after it? How are you doing your diagnostics? Are you in a maintenance policy with the level of maintenance that you're doing. Erm, start to look at the [user] and navigate his report in more detail. So we're gathering more and more data and starting to analyse that data and then coming up with solutions on how we might reduce the [faults]…. And then you get a win-win obviously, because that saves us money and it gives more [asset availability] to the end user. So that, predominantly, is what we aim to do – that support for [users] more than probably the contract would have wanted us to."*

Second, where the customer could do no more, human activities within the firm bridge the gap, albeit with some difficulties:

*"Now you can either spend two years having the fight and whinging or if you have got the relationships you can just, it will get sorted out so….. it just makes everybody's life a lot easier and things just get done."*

Thus, human resources through direct service provision can amplify variety through their responses to the customer so that variety could be absorbed. In other words, human resources create responses that exhibit requisite variety.

Our study also found that human activities were instrumental in co-creating the experience (i.e. emotional value). The firms had to design, within the service system, methods of how individuals' perceptions within the customer organisation were also 'transformed' as part of System 1 operations, i.e. the management of customer experience. The method varied across organisations. One of the firms used technological resources to allow the customer to 'view' the way they worked to create transparency and closeness while the two other firms provided regular updates, even when contractually they didn't need to. All three organisations used relationships so that the customer 'perceived' the contract was in good hands and outcomes were on track.

> *"we're starting to have visual and verbal contact with the people that need to be helping us sort it – so they're starting to become part of it – they're starting to feel it…. it's about us understanding what we're actually delivering and changing our culture, environment, abilities and roles and responsibilities are aligned to it [the customer]"*

> *"I think they trust us; trust us to deliver excellence actually isn't a bad logo for somebody. I think they do trust us; they do know we know what we're talking about. We're excellent at fire-fighting – we're well known for that …. If there's a problem we are the world's best at solving them because that's interesting to us because that's our culture, you-know, we will throw people at issues… And to be quite honest we reward it as well; we reward people for sorting problems out for us."*

*Proposition 5a: Direct service provision ensure viability through absorption of contextual variety and co-creating emotional value and experiences*

Our study also found that contextual variety was a manifestation of latent need, and that the variety of use belies the need for additional provisions from which the firm, if it provided them, could derive greater revenues:

> *"we get into an argument with the [user] that, .... they say 'the outcome isn't what we expected'. Now actually the outcome is what is expected but it's not what they now want because they want more......then what the user wants in terms of [outcome] is more than we've agreed...but it looks like it's going to improve [the] order book position"*

*Proposition 5b: Contextual variety provided an opportunity for firms to innovate and derive new revenues to satisfy customer latent demand*

### 5.2.3 Interaction of Direct and Indirect Service Provision

Our study found that the firm now has to rethink its resources and how System 1 is configured for achieving outcomes, which is drastically different from how it was originally set up to manufacture and transfer ownership of assets.

With the change of System 1 transformation activities from manufacturing to achieving outcomes comes a change in resources required to achieve that co-creation; this in turn comes with the challenge of whether the asset was designed correctly to support such activities. Our study found that an asset designed and engineered for a transfer of ownership to the customer so that the customer achieves the outcomes on their own, may not be the most optimal asset for delivering outcomes together with the customer and where such outcomes could be a responsibility of the firm.

> *"A classic example for me with the [asset], it was designed to be stripped and rebuilt in [our factory]. If we'd done that [at client location] it would have been designed* *differently because we would have taken it apart differently, because [in the factory], we don't have to worry about [shelters to protect the assets] and all those sorts of things ……So there are parameters placed on you which the customer has to deal with in a [use] environment…and you need to now deal with that (when you are delivering outcomes)."*

Our study found that delivering outcomes began with the firm 'wrapping' human activities around an asset, without any serious thought about (a) the outcomes the system aims to achieve; (b) the resource combination of direct and indirect service provision to achieve the same outcomes; and (c) the business model that renders the system viable. Over time, the firms came to the realisation that the asset was not a "sacred cow" and the better it was able to absorb contextual variety of use, the lesser its dependency on embedded human capability and the better it was able to scale and replicate the system across contracts. Concurrently, the firms also became aware that understanding where contextual variety is highest and deploying human activities to absorb variety (either by attenuating or amplifying it) resulted in high satisfaction and the co-creation of emotional and perceptual value and customer experience. This is evidenced by the following quote from one of the employees of the firm when discussing their customer:

*If there's a problem we are the world's best at solving them because that's interesting to us because that's our culture, you-know, we will throw people at issues… I think they do trust us; they do know we know what we're talking about. We're excellent at fire-fighting – we're well known for that ……*

With absorption of variety/co-creating emotional value through human resources and scalability/replicability through assets, the firms started putting in place processes where contextual variety became a conduit for feedback on the degree of substitutability for indirect and direct provision for co-created outcomes, and also to drive both direct and indirect service innovation:

*"As we're starting to collect more data about how the customer uses them, either electronically – so does he know we're getting them? He knows we're getting it but he's happy for us to get that – or via interviews with [users] and those things – it's helping us understand better to look for trends; to look for potential failings of those mechanisms so that we can then, a) stop it happening but also look at that particular area and say, 'well, would we do that differently?"*

*Proposition 6: Scalability and Replicability of Direct Service Provision (people and processes) are dependent on the design of the indirect service provision (asset) for variety absorption*

### 5.2.4 Scalability and Absorptive Resources of the Customer for Value Co-creation

Our study also found that the degree of skills and knowledge for the customer to realise and co-create value interacted directly with both direct and indirect service provisions. Assets which are better platforms for co-creation and which are able to absorb greater variety, either through modularity or clever design, required not only lower skills and knowledge from customer employees but also less of such resources. This implies that the scalability and replicability of the provider system may not merely lie with the firm's direct and indirect service provisions, but with the resources required on the customer side to realise the provisions for outcomes. Conversely, complex assets that had greater technological capabilities required more complex sets of resources to use and operate them. This in turn had an influence on the firm's choice of direct or indirect service provision.

*"if you look at a lot of the land equipment … So to take the average lorry that was used by the Army, it was used … you needed to know how to take engines apart and you'd have to change wheels, you now need almost a degree in Electronics because the whole*

*thing is now computerised so, in a sense, they've actually created a problem there, where at one time running a tank or a lorry was quite cheap, you actually now have to change the type of person who now actually manages that because the average sort-of mechanical person can pick out and can do that – it doesn't get fixed any more……in the past where their Army recruits came in at basic mechanic, 'can you undo that bolt?' they're actually having to come in at graduate level to actually be able to manage and understand the complexity of the equipment they're now getting. "*

Customer resources for co-creation therefore had four types of impact on the firm's service provision. First, the more complex (albeit more technologically advanced) indirect service provision would require more complex customer resources to co-create value. Second, the customer activities to realise and co-create value with the indirect service provision could be more replicable and scalable if the asset was easy to use, providing efficiency gains to the customer. This in turn, meant that the firm's direct service provision became less complex in that the customer required less support. Third, if the asset could absorb greater contextual variety (e.g. a switch for an engine that could allow it to fly through ash), the customer would know what to do in different use situations and less use variety permeates into the firm's system, requiring less direct service provision to absorb the variety. Fourth, customer resources themselves could absorb contextual variety by deploying their own internal resources so that the environment is less disruptive on the system.

*Proposition 7: Customer resource requirement to co-create value in contextual variety changes the nature of direct and indirect service provision by the firm and vice versa*

## 6     Discussion

### 6.1     Value, Variety and Viability - Extending the Service Dominant Logic Approach Towards Organising The Firm

To achieve co-created value-in-use that could be both functional and emotional outcomes, our study found that direct and indirect service provision interacted with customer activities to realise the offerings and the configuration depended on the value to be co-created, contextual variety that needed to be absorbed as well as the need for viability for the provider.

Our findings suggest that four interactions exist in the co-creation system:

Interaction 1: Increasing Scalability & Replicability means redeploying resources to indirect service provisioning
Interaction 2: Increasing Variety absorption and co-creating emotional value and experience means deploying resources to direct service provisioning
Interaction 3: Customer activities that co-create value in contextual variety changes the nature of direct and indirect service provision by the firm and vice versa
Interaction 4: Direct & indirect provision impact on customer resources to co-create value

Our study showed that the difficulty in transforming an organisation may lie not merely in the activities of service personnel, or in processes that surround the asset, but in the design and engineering of the asset itself to support activities of service personnel in combination with customer resources. Consequently, if the asset was originally designed towards a different set of boundaries i.e. the firm is only responsible till the ownership was transferred, it may need to be redesigned with this new set of boundaries where both are now responsible for co-created outcomes.

The firm's value proposition for co-created outcomes consists of both direct (human activities) and indirect (asset) service provision, and the tension between them that threatens viability lies in the degree of replicability and scalability. Our study found that direct service provision challenges the viability of the firm through its inability to scale for growth and replicate across other contracts. The findings indicate that customer-facing teams held the knowledge of the customer, their contexts and their demands within human capability and skills to the extent that although service to the customer was excellent, every contract became a new design, a new team and a new set of relationships. To reduce the threat, firms have to redesign the asset. Yet, we found that direct service provision absorbed contextual variety and co-created emotional outcomes (experiences), leading to higher perceived customer satisfaction. In addition, contextual variety was a manifestation of latent need and new markets and innovation could arise when variety of use is closely monitored.

Our findings suggest a paradox in that as indirect service provision (assets) become more technologically capable and complex which could increase its exchange value to the firm, both the direct service provision (human activities) and the customer resources (resources to co-create value) become less scalable and replicable (and in many cases, more expensive). This in turn could result in an inability in the overall co-creating system to achieve outcomes in a scalable and replicable manner, which may threaten the viability of the firm in the long term. From a business model perspective, the risk of higher co-creating resources by the customer may compel more contracts based on outcomes which could reduce customer co-creating resources but may result in the firm re-engineering the asset for better use capabilities.

### 6.2 A Proposed Viable System of Indirect and Direct Service Provision With Customer Activities for Co-creation

Our study suggests that System 1 of the firm as a viable organisation co-creating functional and emotional outcomes consists of three main System 1 operational elements that interact: That of transforming indirect service provision (materials and equipment), transforming direct service provision (people, information and processes) and transforming the customer employees, as shown in Figure 3. The connections between these System 1 entities are closely coupled, resulting in emergent effects. Serving the three entities are resources accessible by System 2, which consists of a regulatory centre for each element of System 1, and an overseeing regulation at the senior management level. System 2 plays a crucial role in achieving outcomes as it serves not only to regulate the interactions between elements of System 1, but also the most stable and efficient configuration of direct and indirect provision to achieve customer transformation and co-creation within some level of contextual variety. System 2 is therefore tasked with balancing scalability and replicability with variety amplification and attenuation within System 1. To co-create value with customers, System 2 also achieves an important regulatory function. Where the firm is unable to amplify variety to match customer's contextual variety, System 2 has to be able to harness *customer resources* to reduce variety in the system, either through changes of customer use behaviours or through relationships and culture, suggested by Beer as the 'damping of oscillations'.

The viability of a firm transforming from a manufacturing concern into a service organisation co-creating valued outcomes therefore, concretely implies:

1. The redrawing of system boundaries to include the customer within its boundaries but which *must also include* Systems 3 and 2's capability to harness customer resources to amplify or attenuate variety in the system caused by uncertain environmental factors;
2. The additional System 1 element that *transforms customer employees for perceptual and emotional value and customer experience* in addition to transforming indirect service provision (design and manufacturing of asset) and direct service provision (design and implementation of people and processes);

3. The customer transformation operational element could be interventionistic on the customer's co-creating activities at a lower level of recursion and which the firm may not have control over;
4. A more tightly coupled System 1 operational entities where transforming indirect service provision (design and manufacturing of asset) for value co-creation with the customer interacts with transforming direct service provision (design and implementation of people and processes) as well as with customer co-creation activities. A tightly coupled System 1 creates emergent effects embedded within the customer experience;
5. System 2's ability to coordinate between the three operational entities through allocation of different resources required for scalability/replicability and variety amplification/attenuation through redesign of direct or indirect service provision over time; and
6. The support from Systems 3, 4 and 5 to allocate resources and control the overall system.

# 7 Conclusion

Beer's first axiom of management suggest that the sum of horizontal variety disposed by all the operational elements must be equal to the sum of vertical variety disposed by the six vertical components of corporate cohesion. Our study suggests that organisations structured around manufacturing require a re-evaluation of operational elements and viability within the system when they transform towards being a full service organisation. Homeostasis could be seriously disrupted if they are not able to do so and the viability of the system would be threatened. We propose that the understanding of value-in-use (including all practical, logical, experiential and emotional value), contextual variety, and a system's perspective of viability are the **three core principles** for designing an organisation that is able to co-create value with customers through both direct and indirect service provision.

Our study extends the work in S-D Logic. Specifically, operand and operant resources in the context of value co-creation is formed from direct and indirect service provision of the firm together with customer activities to realise the offerings in context. Our work provides greater understanding of value co-creation in a complex system that includes the discussion on the firm's viability as it invests in such a capability.

More importantly, our work contributes to the understanding of the interface between equipment (goods) and human activity as direct and indirect service provision for co-creating value with customers. Goods are often designed purely within the domain of engineering and product design, with the combination of human activity and goods often placing human activity as a supporting role to the equipment. This study considers the design of both equipment and human activity on an equal footing for value co-creation with the customer, and it yielded interesting results on when direct provisioning (goods) should be redesigned, considering all activities equally.

**References**

Arnould, E. and Thompson, C. (2005), "Consumer culture theory (CCT): Twenty years of re- search", *Journal of Consumer Research* Vol 31, pp. 868–882.

Ashby, W. (1956), *An Introduction To Cybernetics*, Taylor & Francis,

Ashby, W. (1958), "Requisite variety and its applications for the control of complex systems", *Cybernetica*, Vol 1, pp. 53–99.

Ashby, W. (1969), *Self-Regulation And Requisite Variety,* in F.Emery (Ed.), *Systems Thinking*, Penguin Books, Harmondsworth, pp. 105–124.

Baines, T.S., H W Lightfoot, S Evans, A Neely, R Greenough, J Peppard, R Roy, E Shehab, A Braganza, A Tiwari, J R Alcock, J P angus, M Bastl, A Cousens, P Irving, M Johnson, J Kingston, H Lockett, V Martinez, P Michele, D Tranfield, I M Walton and H Wilson (2007) '*State-of-the-art in product-service systems.*' *Proceedings of the Institution of Mechanical Engineers, Part B: Journal of Engineering Manufacture*, Vol. 221, No.10

Ballantyne, D. and Varey R. (2006), "Creating value-in-use through marketing interaction: The exchange logic of relating, communicating and knowing", *Marketing Theory,* Vol. 6 No. 3, pp. 335.

Beer, S. (1979), *The Heart Of Enterprise,* Wiley, New York.

Beer, S. (1981), *Brain of the Firm*, John Wiley & Sons

Beer, S., (1984)"The Viable System Model: Its Provenance, Development, Methodology and Pathology", Journal of the Operational Research Society, Vol.35 No.1, pp. 7-25.

Beer, S. (1985), *Diagnosing The System For Organizations,* Wiley

Belk, R. and Sherry Jr, J. (Eds) (2007), *Consumer Culture Theory: Research in Consumer Behavior,* Elsevier, New York.

Bengtsson, D. (2004), Pleasure and the phenomenology of value, in Rabinowicz, W. and Rönnow-Rasmussen, T. (Eds.), *Patterns of value –Essays on Formal Axiology and Value Analysis Vol 2,* Lund Philosophy Reports.

Beverland M., Farrelly F. and Woodhatch Z. (2004), "The role of value change management in relationship dissolution: Hygiene and motivational factors", *Journal of Marketing Management* Vol. 20 No. 9, pp. 927–939.

Boothroyd, G. (1994), "Product design for manufacture and assembly," *Computer-Aided Design*, Vol. 26, No. 7, pp. 505–520.

Cook, P. and Graham, D. (1977), "The demand for insurance and protection: The case of irreplaceable commodities," *The Quarterly Journal of Economics,* Vol. 91, No. 1, pp.143–156.

Cutland, N. (1980), *Computability: An Introduction To Recursive Function Theory*. Cambridge University Press, Cambridge.

Dooley, D. (2001) *Social Research Methods*. Prentice-Hall, Inc

Flint, D., Woodruff, R., Gardial, S. (2002), "Exploring the phenomenon of customers' desired value change in a business-to-business context", *The Journal of Marketing* Vol. 66, No. 4, pp.102–117.

Forrester, J. (1968) "Market growth as influenced by capital investment", *Industrial Management Review* Vol. 9, No. 2, pp.83-105.

Godsiff, P. (2000), "Service systems and requisite variety", in *The 2009 Naples Forum on Service: Service-Dominant Logic, Service Science and Network Theory*, pp. 978–88

Golinelli, G. (2010) *Viable Systems Approach (VSA): Governing Business Dynamics*, CEDAM

Grönroos, C. and Ravald, A. (2010), "Service as business logic: Implications for value creation and marketing", *Journal of Service Management*, Vol. 22, No. 1, pp. 1–1.

Haglund, B. (1988), "Report on discrimination powers for hartmann-type value hierarchies", Tech. rep., Department of Philosophy, University of Gothenburg

Hartman, R. (1973), *The Hartman Value Profile (Hvp): Manual Of Interpretation,* Research Concepts, Muskegon, MI

Hartman, R., Plochmann, G., and Weiss, P. (1967), The *Structure Of Value: Foundations Of Scientific Axiology*, Southern Illinois University Press Carbondale, IL

Heinonen, K. and Strandvik, T. (2009), "Monitoring value-in-use of e-service", *Journal of Service Management,* Vol. 20, No. 1, pp. 33–51.

Heylighen, F. and Joslyn, C. (2001), "Cybernetics and second order cybernetics", *Encyclopaedia Of Physical Science & Technology* Vol. 4, pp. 155–170.

Husserl, E., Landgrebe, L., Churchill, J. and Ameriks, K. (1973), *Experience And Judgment: Investigations In A Genealogy Of Logic,* Routledge, London.

Jessop, B. (2003), "Governance and metagovernance: On reflexivity, requisite variety, and requisite irony", in Bang, H.P. (Ed.), *Governance As Social And Political Communication,* Manchester University Press, Manchester, pp. 101–16.

Johnstone, S., Dainty, A. and Wilkinson, A. (2009), "Integrating products and services through life: An aerospace experience", *International Journal of Operations & Production Management*, Vol. 29, No. 5, pp. 520–538.

Karni, E. (1983), "Risk aversion for state-dependent utility functions: Measurement and applications", *International Economic Review,* Vol. 24, No. 3, pp. 637–647.

Lapierre, J., Tran-Khanh, A. and Skelling, J. (2008), "Antecedents of customers' desired value change in a business-to-business context: Theoretical model and empirical assessment", *Services Marketing Quarterly* Vol. 29, No. 3, pp. 114–148.

Lemon, K., White, T. and Winer, R. (2002), "Dynamic customer relationship management: Incorporating future considerations into the service retention decision", *Journal of Marketing* Vol. 66, No. 1, pp. 1–14.

Macdonald, E., Martinez, V. and Wilson, H. (2009), "Towards the assessment of the value-in- use of product-service systems: A review," in: Performance Management Association Conference, New Zealand

Marx, K. (1867), *Capital: A critique of political economy (i): The process of capitalist production*, History of Economic Thought Books 1.

Mattsson, J. (1992), "A service quality model based on an ideal value standard," *International Journal of Service Industry Management*, Vol. 3, No. 3, pp.18–33.

Moore, G. and Baldwin, T. (1993), *Principia Ethica*. Cambridge University Press

Neely, A. (2008), Exploring the financial consequences of the servitization of manufacturing. Operations Management Research 1(2):103–118

Ng, I. (2008), *The Pricing and Revenue Management of Services, Advances in Business and Management Studies series*, Routledge, London.

Ng, I., Maull, R. and Yip, N. (2009), "Outcome-based contracts as a driver for systems thinking and service-dominant logic in service science: Evidence from the defence industry", *European Management Journal,* Vol. 27, No. 6, pp. 377–387.

Ng, I., Parry, G., Smith, L., Maull, R. (2010), "Value co-creation in complex engineering service systems: Conceptual foundations," University of Exeter Department of Management Discussion Paper Series 10/04

Ng, I., Guo, L. and Ding, Y. (2011), "The use of information technology as value co-creation: The role of contextual use variety and means drivenness", under review for the special issue of *MISQuarterly on Service Innovation in the Digital Age*

Oliva, R. and Kallenberg, R. (2003), "Managing the transition from products to services", *International Journal of Service Industry Management*, Vol. 14, No. 2, pp. 160–172.

Palmatier, R. (2008), "Interfirm relational drivers of customer value", *Journal of Marketing,* Vol. 72, No. 4, pp. 76–89.

Pawar, K., Beltagui, A., and Riedel, J. (2009), "The PSO triangle: designing product, service and organisation to create value," *International Journal of Operations & Production Management,* Vol. 29, No. 5, pp. 468–493.

Payne, A., Storbacka. K. and Frow, P. (2008), "Managing the co-creation of value". *Journal of the Academy of Marketing Science,* Vol. 36, No. 1, pp. 83–96.

Plato (360 B.C.E), The Republic

Png, I. (1989), "Reservations: Customer insurance in the marketing of capacity", *Marketing Science,* Vol. 8, No. 3, pp.248–264.

Potts, C. and Hsi, I. (1997), "Abstraction and context in requirements engineering: Toward a synthesis"; *Annals of Software Engineering*, Vol. 3, No. 1, pp. 23–61.

Richardson G. and Pugh, A. (1981), *Introduction To System Dynamics Modeling With DYNAMO,*" MIT Press, Cambridge, MA, USA.

Sandström S, Edvardsson B, Kristensson P, Magnusson P (2008) Value in use through service experience. *Managing Service Quality,* 18(2):112–126

Shugan S, Xie J (2000), "Advance pricing of services and other implications of separating purchase and consumption", *Journal of Service Research,* Vol. 2, No. 3, pp.227.

Tuli, K., Kohli, A. and Bharadwaj, S. (2007), "Rethinking customer solutions: from product bundles to relational processes", *Journal of Marketing,* Vol. 71, No. 3, pp. 1–17.

Vandermerwe, S. and Rada, J. (1988), "Servitization of business: Adding value by adding services", *European Management Journal,* Vol. 6, No. 4, pp. 314–324.

Vargo, S. and Akaka, M. (2009), "Service-dominant logic as a foundation for service science: clarifications", *Service Science,* Vol. 1, No. 1, pp. 32–41.

Vargo, S. and Lusch, R. (2004), "Evolving to a new dominant logic for marketing," *The Journal of Marketing,* Vol. 68, No. 1, pp. 1–17.

Vargo, S. and Lusch, R. (2008), "Service-dominant logic: Continuing the evolution", *Journal of the Academy of Marketing Science,* Vol. 36, No. 1, pp. 1–10.

Vargo, S., Maglio, P. and Akaka, M. (2008), "On value and value co-creation: A service systems and service logic perspective," *European Management Journal,* Vol. 26, No. 3, pp. 145–152.

Weber, M. (1909), "Zur methodik sozialpsychologischer enqueten und ihrer bearbeitung", *Archiv fur Sozialwissenschaft und Sozialpolitik,* Vol. 29, No. 3, pp. 949–958.

Xie, J. and Shugan, S. (2001), "Electronic tickets, smart cards, and online prepayments: When and how to advance sell", *Marketing Science,* pp. 219–243.

Yin, R. (2003), *Case Study Research: Design And Methods,* Sage Publications, Inc,

**Figure 1:** A Viable System Model (source: Beer, 1984)

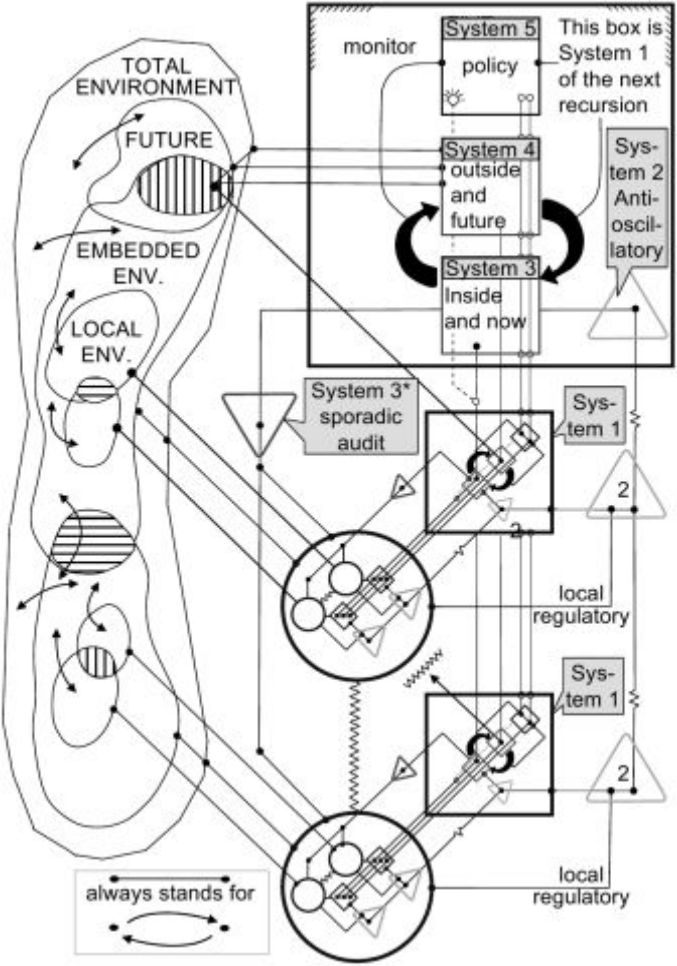

**Figure 2:** Interactions Between Customer Resources and Activities and the Firm's Direct and Indirect Service Provision in a System of Value Co-creation

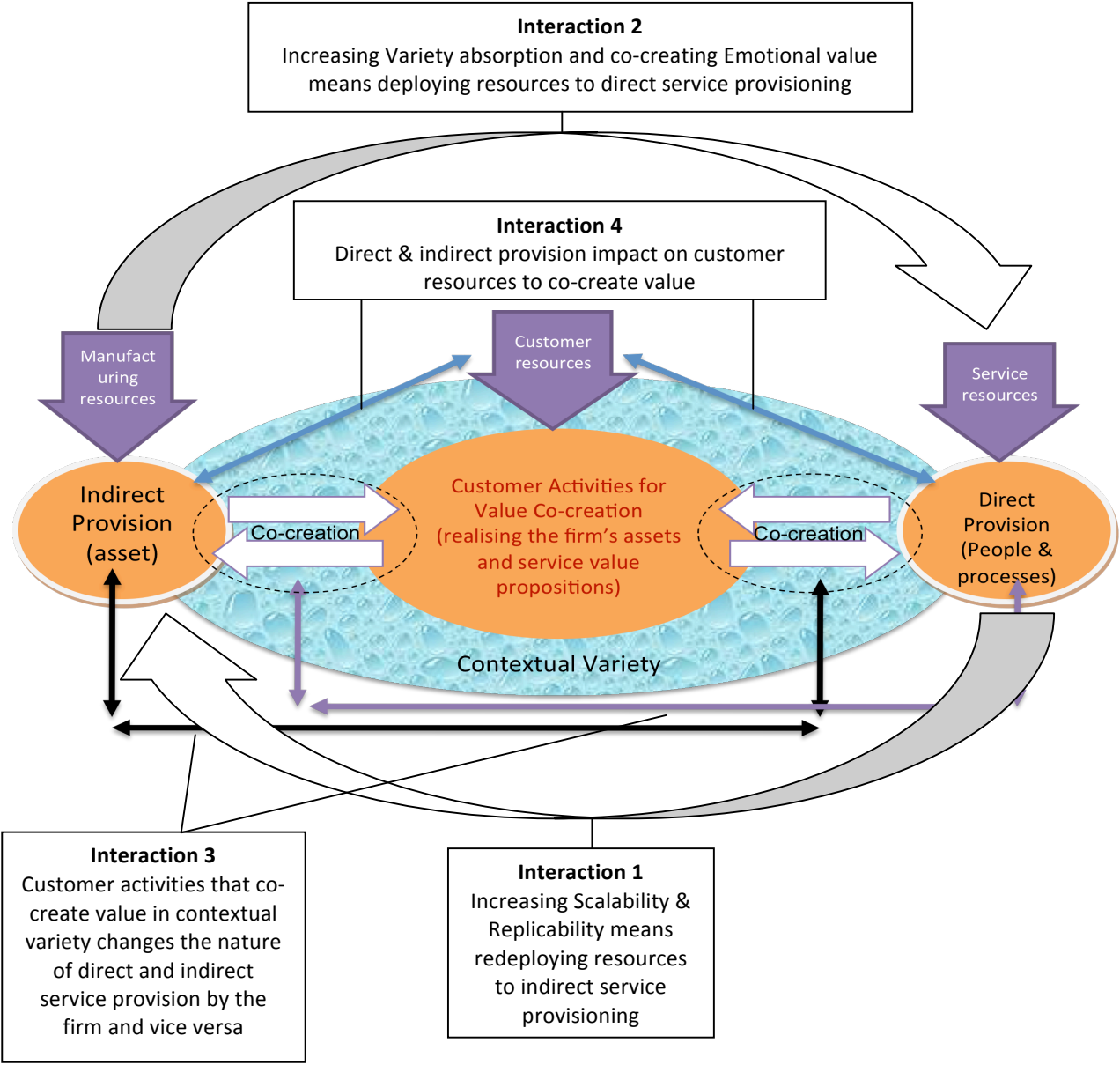

**Figure 3:** A Viable System for an Organisation Co-Creating Outcome-Based Value In Use

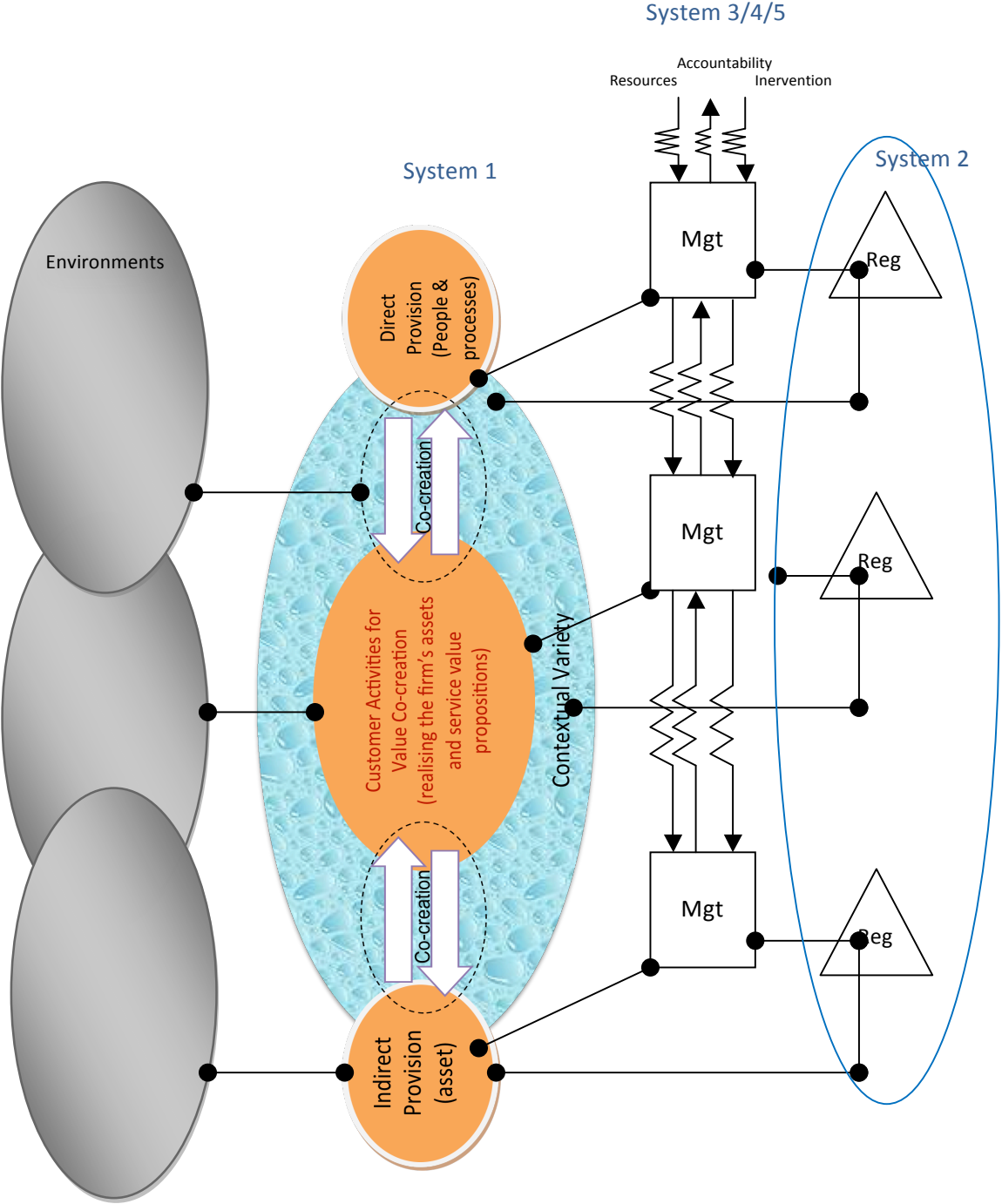

**Table 1:** Beer's Viable Systems Model

| SYSTEM | Description | Elaboration | Traditional company functions | Human body functions |
|---|---|---|---|---|
| 1 | **Key transformation** | This system has to deliver what it has been designed to do, despite changes in the environment, so it must have the capacity to adapt to be able to cope and return the entity to stability. A firm's environment consists of its customers, suppliers, regulators, all of which could experience perturbation which could disrupt the firm's core operations. | Operations Management – core value transformations. Recursions of viable systems | All the muscles and organs. The parts that actually DO something. The basic activities of the system. |
| 2 | **Conflict resolution, stability, coordination** | System 2 coordinates between the various recursions in System 1, so that common functions could be coordinated within the group efficiently. Note that System 2 is not autonomous, as none of the activities earn any revenues, although having an effective System 2 could save costs for the firm. | Account payable/receivable IT support Health and Safety Travel Tax Compliance Administration | The sympathetic nervous system which monitors the muscles and organs and ensures that their interactions are kept stable. |
| 3 | **Internal regulation, Optimisation, Synergy.** | System 3 is the executive function of the group. The firm should be organised in such a way that the whole firm benefits, and even though some parts of the firm may not have the direct incentive to operate for the collective, System 3 ensures that they do, often leading to resource bargaining and lobbying. System 3 star is the part of System 3 that is required occasionally to enter System 1, often to cope with a crisis. System 3 star often includes internal audit, finance audit or compatibility audit where the purpose is not to micro-manage but to do a check to ensure System 1's effectiveness and agility. | Management accounting, production control. operations planning and control /audit – rules, resources, rights, responsibilities – interface between 4/5 and 1/2 | The Base Brain which oversees the entire complex of muscles and organs and optimises the internal environment. |
| | **Adaptation, dealing with a changing environment, forward planning.** | System 4's role is to scan the horizon, observe and forecast a future and plan for it. To do so, it must have a clear view of System 3 (current state) and where it needs to go to ensure survival. System 4 has ongoing conversations between its current state and its future state, setting up future resources and developing new offerings. Systems 3/4 homeostat is expected to maintain the tension between a future state and the current state. | Management, marketing, strategy, environment scanning (for adaptability) | The Mid Brain. The connection to the outside world through the senses. Future planning. Projections. Forecasting. |
| 5 | **Ultimate authority, policy, ground rules, identity.** | System 5's job is to maintain the System 3/4 homeostat, ensuring that the firm survives at present and remain viable for the future. System 5 also tackles the issue of the firm's identity and its mission. Much of business policy and strategic governance sits within System 5, which asks if the firm is doing the 'right' thing, rather than just doing it right. System 5 also manages the vertical variety of its own system from System 1 to 5, while balancing the horizontal variety between the systems and the environment. | Board of directors, business policy (decisions to maintain entity, balance demands from all parts, steer the organisation) | Higher brain functions. Formulation of Policy decisions. Identity. |